\documentclass[sigconf,nonacm]{acmart}
\usepackage{array}

\AtBeginDocument{%
  }

\copyrightyear{2023}
\acmYear{2023}

\begin{document}

\title{3Description}
\subtitle{An Intuitive Human-AI Collaborative 3D Modeling Approach}

\author{Zhuodi Cai}
\orcid{0009-0000-1023-1935}
\affiliation{
  \institution{Tisch School of the Arts, New York University}
  \state{New York}
  \country{USA}
}
\thanks{This is the author’s accepted manuscript. The final version was published in \textit{ARTECH '23: Proceedings of the 11th International Conference on Digital and Interactive Arts}, ACM, 2024. DOI: \url{https://doi.org/10.1145/3632776.3632785}. This work is licensed under CC BY-NC-ND 4.0. Attribution required. Non-commercial use only. No derivatives.}


\begin{abstract}
This paper presents 3Description, an experimental human-AI collaborative approach for intuitive 3D modeling. 3Description aims to address accessibility and usability challenges in traditional 3D modeling by enabling non-professional individuals to co-create 3D models using verbal and gesture descriptions. Through a combination of qualitative research, product analysis, and user testing, 3Description integrates AI technologies such as Natural Language Processing and Computer Vision, powered by OpenAI and MediaPipe. Recognizing the web has wide cross-platform capabilities, 3Description is web-based, allowing users to describe the desired model and subsequently adjust its components using verbal and gestural inputs. In the era of AI and emerging media, 3Description not only contributes to a more inclusive and user-friendly design process, empowering more people to participate in the construction of the future 3D world, but also strives to increase human engagement in co-creation with AI, thereby avoiding undue surrender to technology and preserving human creativity.
\end{abstract}


\begin{teaserfigure}
    \centering
  \includegraphics[width=17.5cm]{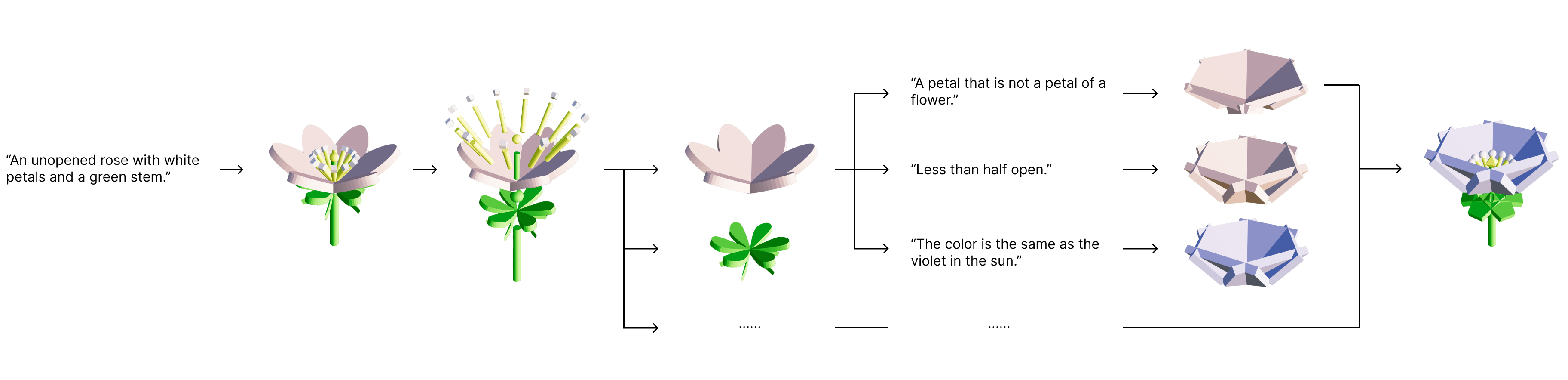}
  \caption{3Description allows users to generate a model and adjust its components by voice input. The process consists of three stages including (1) model generation, (2) component segmentation, and (3) component modification.}
  \label{fig:teaser}
\end{teaserfigure}


\maketitle

\section{Introduction}
In today's rapidly evolving technology landscape, 3D modeling has become crucial in various fields such as entertainment, education, healthcare, and architecture. The market for 3D modeling is expanding rapidly, with key players witnessing significant growth in user numbers. Blender reported a 3.63 million increase in new users and an additional 4.73 million downloads in 2020 compared to the previous year\cite{Blender21Siddi}. Furthermore, there is a growing social significance associated with 3D creation. One notable example is Metamorphic\cite{metamorphic}, a VR social experience introduced in 2020. This immersive platform allows multiplayer to influence various attributes of the 3D drawings through movement and interactions, thereby reshaping the 3D environment.

However, traditional 3D modeling tools present challenges in terms of accessibility and usability for non-professionals. Learning 3D modeling from scratch requires substantial time and effort, particularly for individuals without artistic or design backgrounds who may struggle with it. Additionally, approximately 10\% of the global population experiences hand and wrist pain\cite{ferguson2019wrist}, further complicating the use of conventional mouse-based modeling tools. These limitations restrict broader participation in the creation and exploration of the 3D world.

\section{Background}
Since the introduction of Sketchpad\cite{sutherland1963sketchpad} in 1963, which is considered the earliest milestone in computer-aided design, the research and industrial communities have collaboratively driven the application of computer graphics, including 3D modeling, for over 60 years. Their efforts have focused on areas like enhancing functionality, optimizing operational efficiency, and improving usability.

To deal with the challenge of intuitive 3D modeling, traditional professional software like Autodesk AutoCAD 2000\cite{autocad2000} started integrating voice commands through plugins like Voice Xpress 20 years ago. Today, with the release of GPT-4 in March 2023, BlenderGPT has been developed and connected to Blender, enabling more flexible text command inputs. Alongside these command operation innovations, products such as Shapr3D\cite{shapr3d}, which facilitates modeling on the iPad with fingers, and Gravity Sketch\cite{gravitysketch}, a VR tool that allows drawing in the air with controllers, have emerged.

While these tools offer simplified interfaces and enhanced interactive capabilities, questioning the need for a mouse and providing more interactive models, they still assume users possess drawing skills and modeling knowledge, making them less accessible to individuals without an artistic or design background. This limitation becomes more apparent when the tool requires users to start from a blank canvas and create designs from scratch.

To provide those without drawing skills with greater opportunities to create, AI technologies are playing an increasingly significant role. In the field of 3D generation, there have been various studies that leverage text prompts to generate a range of 3D forms. This includes POINT-E\cite{nichol2022point} for point cloud models, Dream Fields\cite{jain2022zero} for mesh models, and CLIP-Forge\cite{sanghi2022clip} for voxel models. 

However, for non-professionals lacking an artistic or design background, the process of generating a plausible result based on a textual description is merely the first step. To avoid relying solely on AI-generated outcomes and to ensure that the generated results align with the user's original idea, subsequent modifications are necessary. This still presents barriers in terms of using specialized terminology and effectively communicating specific design requirements.

\section{Methodology}
To investigate how individuals without an art or design background describe the appearance of 3D objects, particularly focusing on 3D flowers, a qualitative research approach was adopted. The individual interviews were conducted with students from 24 to 27 years old who do not have an art or design background (N = 9).

During the interviews, participants were instructed to imagine the appearance of a 3D flower and describe it using any means or tools without limitations. The researcher observed and recorded participants' verbal descriptions and accompanying gestures on paper. Initially, participants provided very basic descriptions, which were further elicited through probing questions. Hand gestures were commonly used to supplement verbal descriptions, representing the shape and positional relationships between components. Afterward, participants were encouraged to draw their imagined 3D flower on paper using a pen, despite initial hesitancy and lack of confidence expressed by everyone. These drawings were later compared to verbal and gestural descriptions to identify any discrepancies.

\section{Findings}
The study with individuals from non-art or design backgrounds yielded the following significant findings:

\subsection{Initial descriptions}

Before probing, participants provided rough descriptions that primarily covered basic visual features, such as an open pink flower with a green stem.

\subsection{Probing for details}
Through probing questions, interviewees were encouraged to provide more specific details about the 3D flower's appearance. Consequently, they began incorporating additional information, such as the round shape of the petals, the similarity between the leaves and petals, the flower's opening angle of 60 degrees, and the light yellow color of the flower's center. Participants frequently used gestures to assist in describing shapes and positioning, with hand movements being the preferred method.

\subsection{Use of analogies}

Participants commonly employed analogies with reference objects to describe shapes and colors. For example, "The petals look like rose petals" and "The color is the purple of the eggplant skin."

\subsection{Difficulty with numeric descriptions}

Participants found it challenging to verbally describe numeric aspects of the 3D flower, such as angles and proportions unless the values were extreme and easy to conceptualize. They relied more on gestures and visual comparisons rather than precise numerical descriptions.

\subsection{Discrepancies in drawings and descriptions}
It was observed that the strokes were not smooth but intermittent in the drawings. Besides, not only was it difficult to depict complex structures in detail as previously described, but discrepancies also occurred in basic shape, position, and size. For example, one participant mentioned that she lacked drawing skills and unintentionally drew rose petals in the shape of lotus petals.

\subsection{Visualization preferences}

In the final feedback session of the interview, participants expressed a strong desire to visualize the described 3D flower. For example, one interviewee noted she prefer being provided with related visualized images for better communication, while another expressed interest in having someone create several drawings based on his description to let him assess satisfaction. One participant mentioned he was unfamiliar with specialized terminology, so he had no choice but to use analogy as a way of visualization. This indicates that non-professionals prefer not to start with a blank canvas and instead require some form of visualization for potential solutions.

\section{Concept}

According to the findings, it is needed to design an intuitive interaction method that is more in line with the user's habit of communicating 3D appearance ideas in a natural state, giving real-time visual feedback. To guide the user to provide more details of the model, once users describe the 3D appearance they have in mind, it is necessary to display not only the overall model's appearance but also separable model components with corresponding position constraints. This allows for detailed adjustments in shape, position, color, and other aspects. For example, a flower can be broken down into components such as petals, sepals, and a receptacle. An angle constraint exists between the flower's petals and the receptacle, which can be controlled to open or close the petals. Users can express their ideas through voice input, generating a prototype model with the mentioned characteristics. Based on visual feedback, users can then provide more specific verbal descriptions for each component and make further adjustments using gesture recognition. The extent of separability of the model components determines the granularity of adjustments. For instance, the stamen of a flower can be further divided into anther and filament, and the anther can be separated into pollen sacs, pollen grains, and a line of dehiscence.

\begin{figure}[h]
  \centering
  \includegraphics[width=\linewidth]{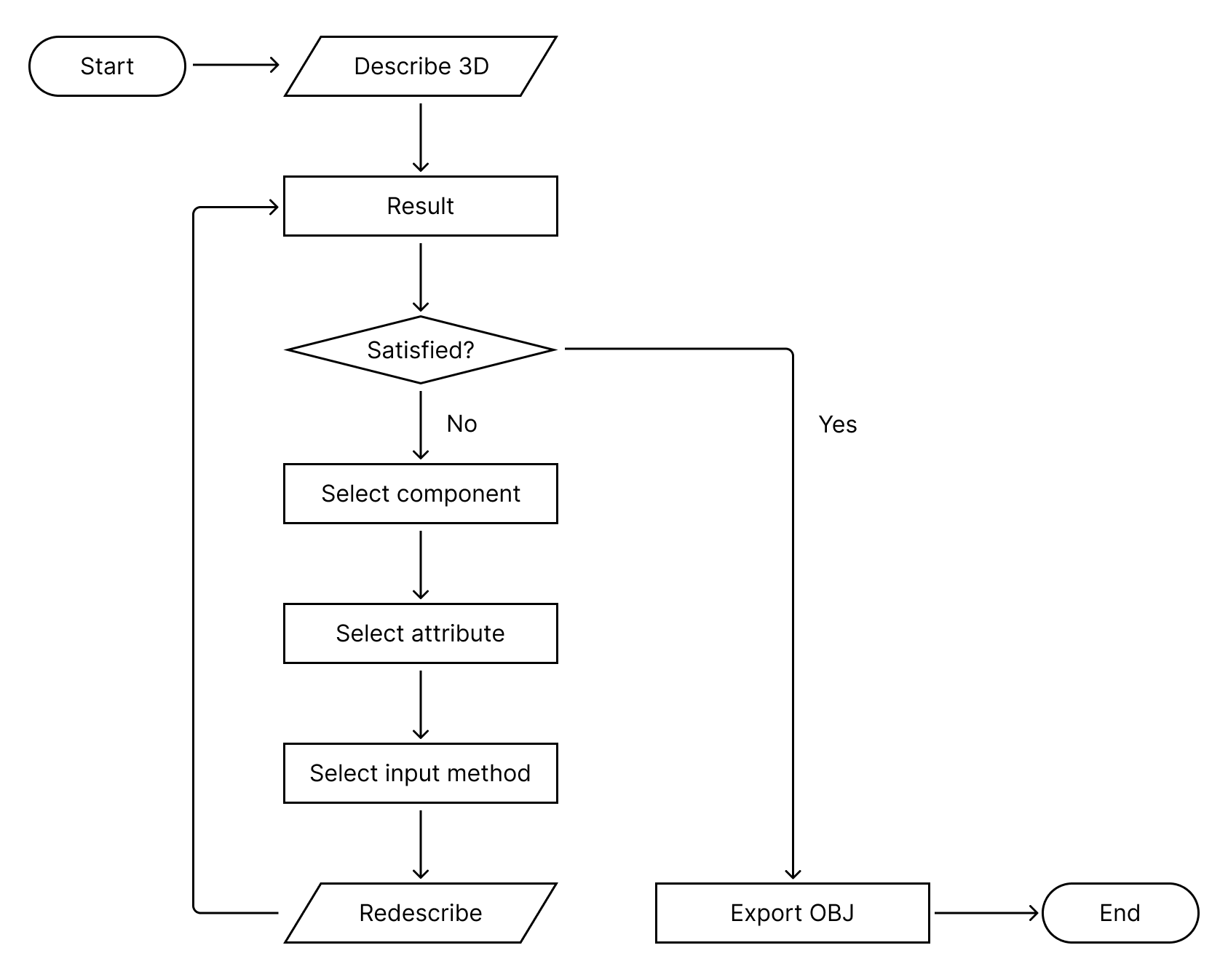}
  \caption{3Description user flow}
\end{figure}

\section{Implementation}

3Description is developed with JavaScript, integrating OpenAI\cite{brockman2016openai} and MediaPipe\cite{lugaresi2019mediapipe} APIs for simulating dialogues with users and implementing gesture recognition. The generated model code is adapted to the format of Three.js\cite{threejs} and can be compiled seamlessly. API calls are made using Axios\cite{axios}. Leveraging the benefits of web applications, 3Description only requires a camera-equipped mobile device for immediate use, without additional equipment or configuration. The specific implementation of dialog and gesture recognition will be described in the next subsection.

\subsection{Model generation with voice}

In order to serve users without expertise in 3D modeling, 3Description doesn't require standard specialized modeling language inputs from users. Instead, it can interpret the user's abstract and imprecise descriptions, extract relevant keywords, and construct models that fulfill the user's requirements.

The implementation relies on ChatGPT\cite{openai2021chatgpt} GPT-3.5 Turbo and Whisper\cite{hinton2012deep} Whisper-1 from OpenAI. Users are simply required to express their thoughts through speech. Upon receiving the user's speech data, 3Description utilizes the Whisper model to convert the voice into textual content. Subsequently, it combines the text data with a specific prompt to query ChatGPT and retrieve a code block in the Three.js format. Finally, 3Description searches the current code using a customized regular expression, replaces it with the generated code block and recompiles it. This approach enables the dynamic generation and rendering of models according to the user's requirements. Depending on the network and frequency of requests, the response time of 3Description typically falls within a few seconds.

\begin{table}
   \caption{Shape generation and modification}
  \label{tab:image}
  \begin{tabular}{>{\centering\arraybackslash}m{0.8in} *3{>{\centering\arraybackslash}m{0.65in}} @{}m{0pt}@{}}
    \toprule
    I/O & Case 1 & Case 2 & Case 3\\
    \midrule
    Voice Input & "Rectangle." & "Similar to rose petal." & "Looks like my soul." \\
    Voice Output & \includegraphics[width=1.5cm]{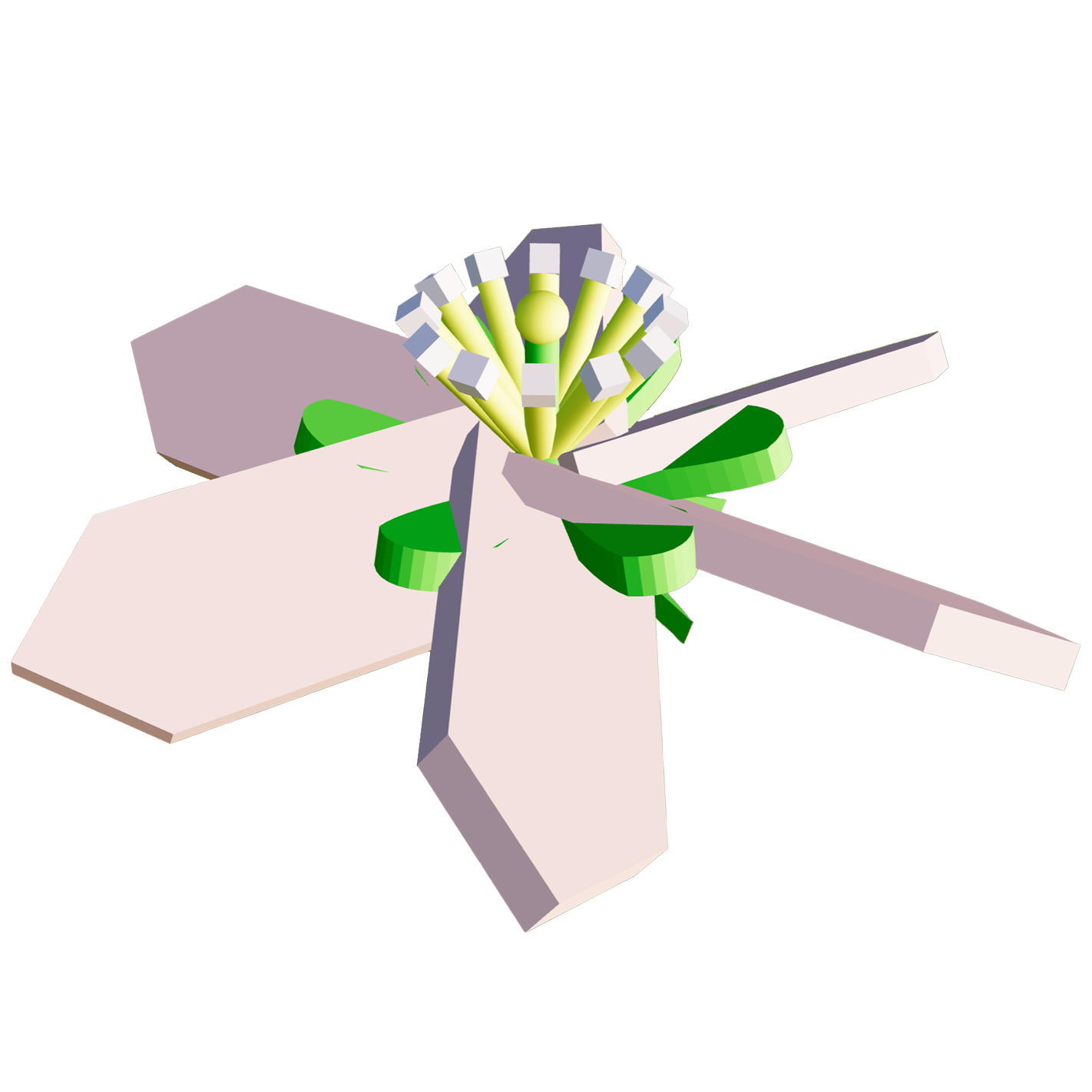}& \includegraphics[width=1.5cm]{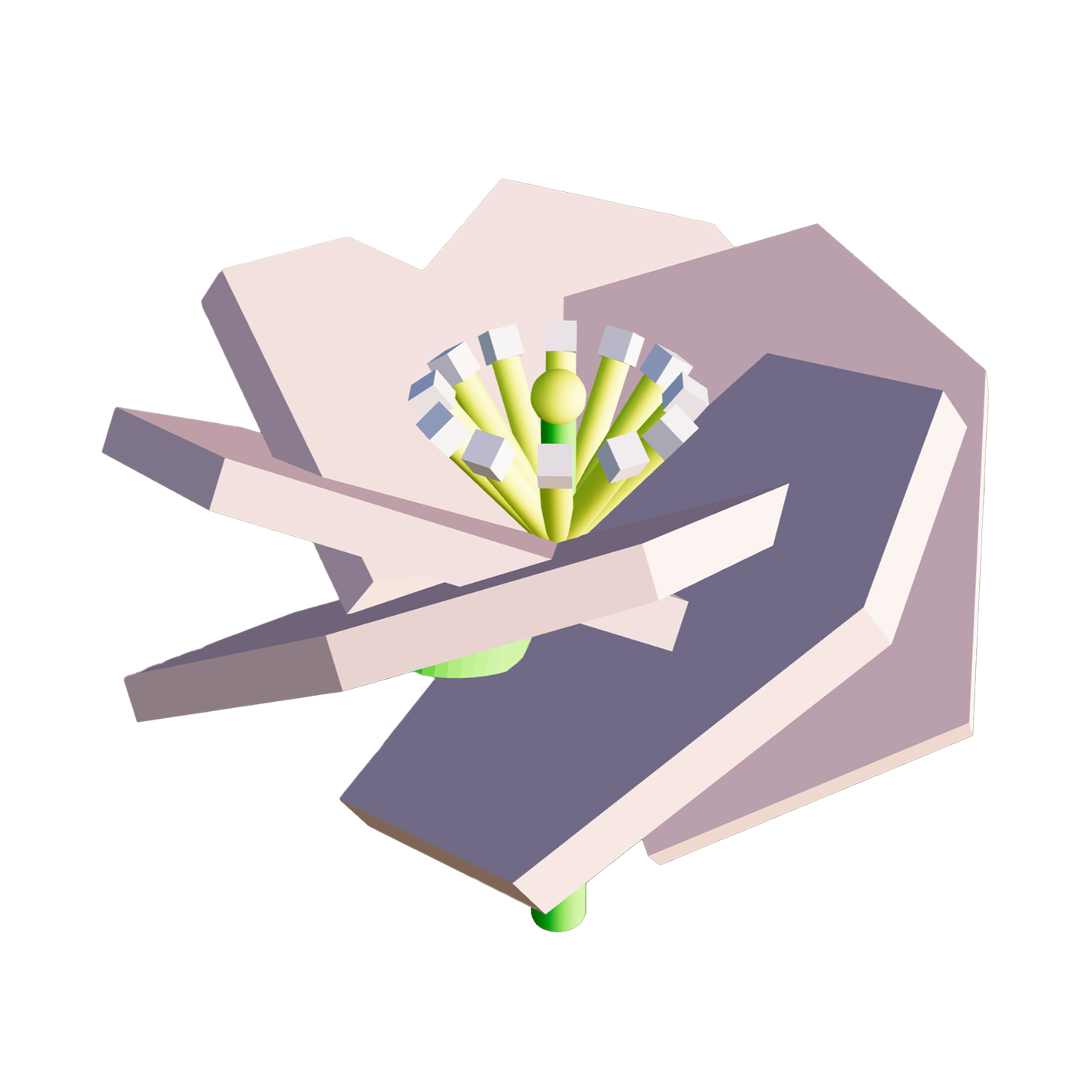}& \includegraphics[width=1.5cm]{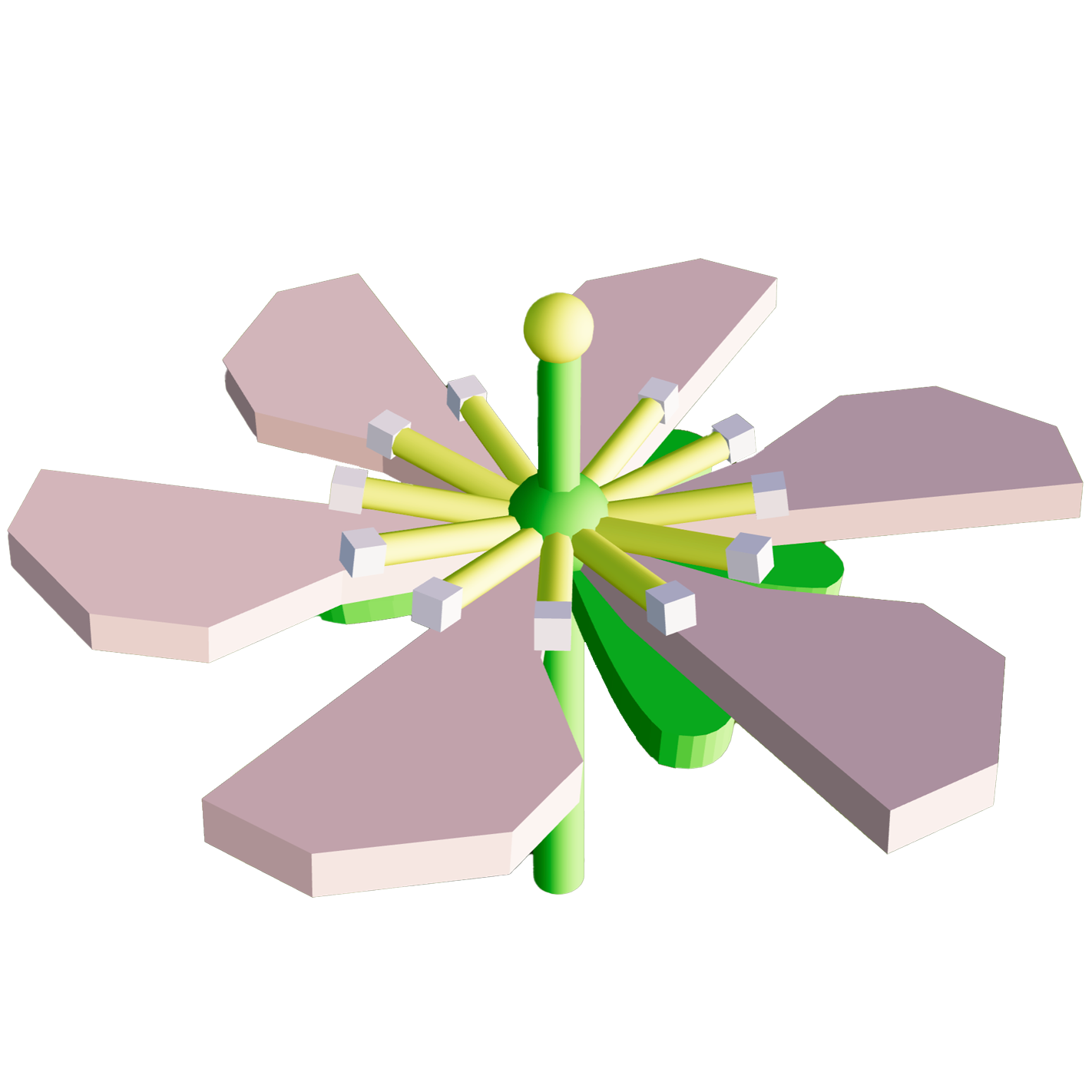} \\
    Gesture Input& \includegraphics[width=1.2cm]{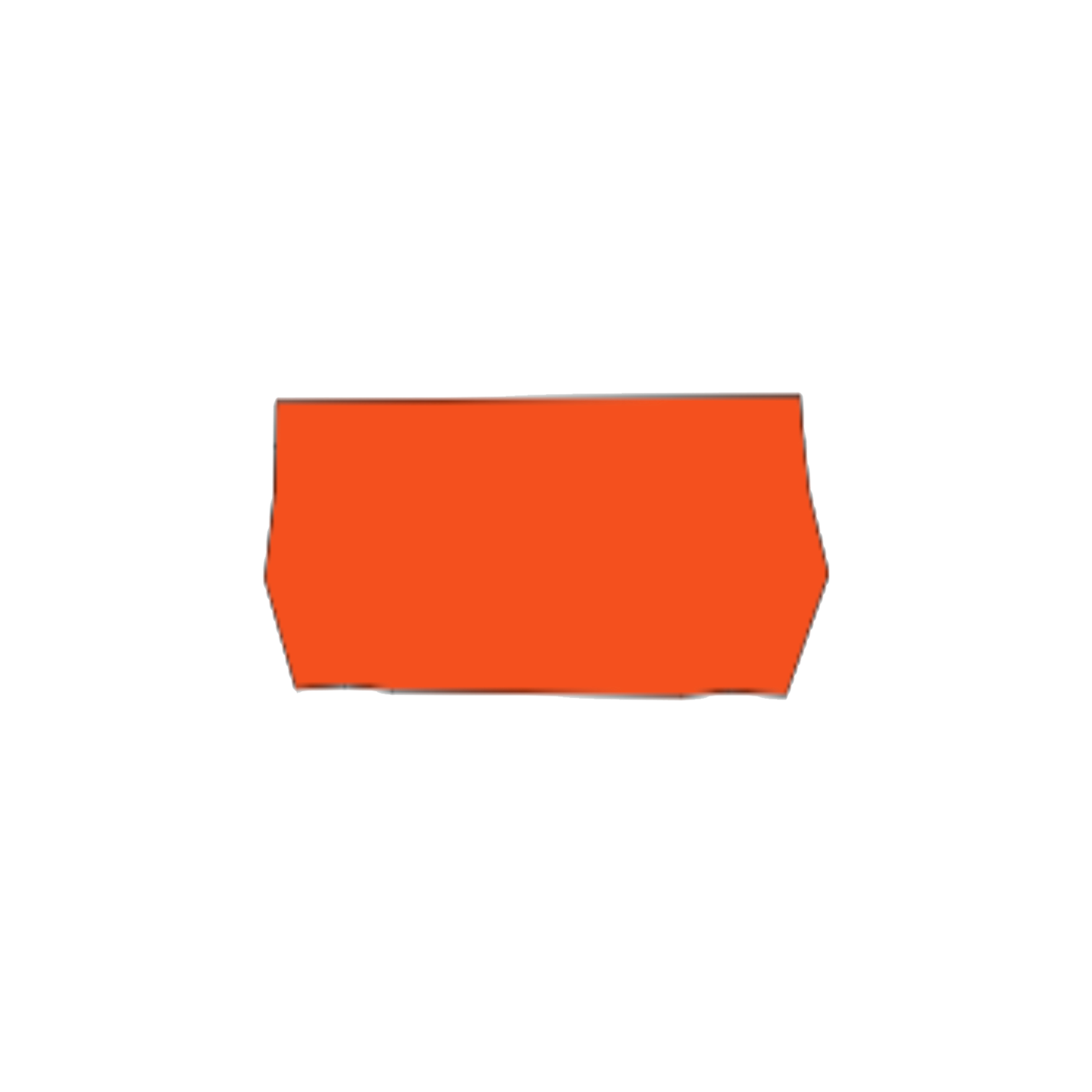}& \includegraphics[width=1.2cm]{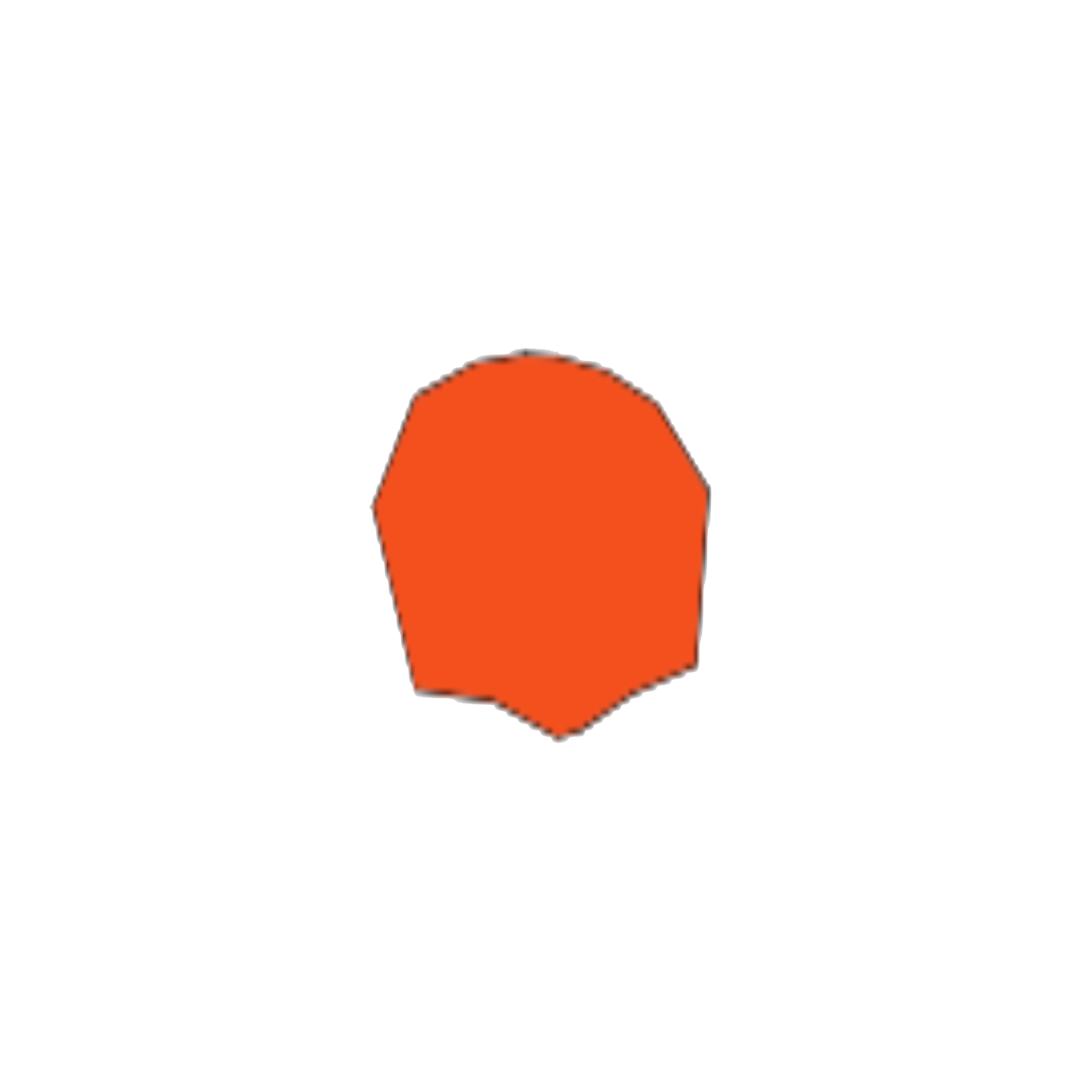}& \includegraphics[width=1.2cm]{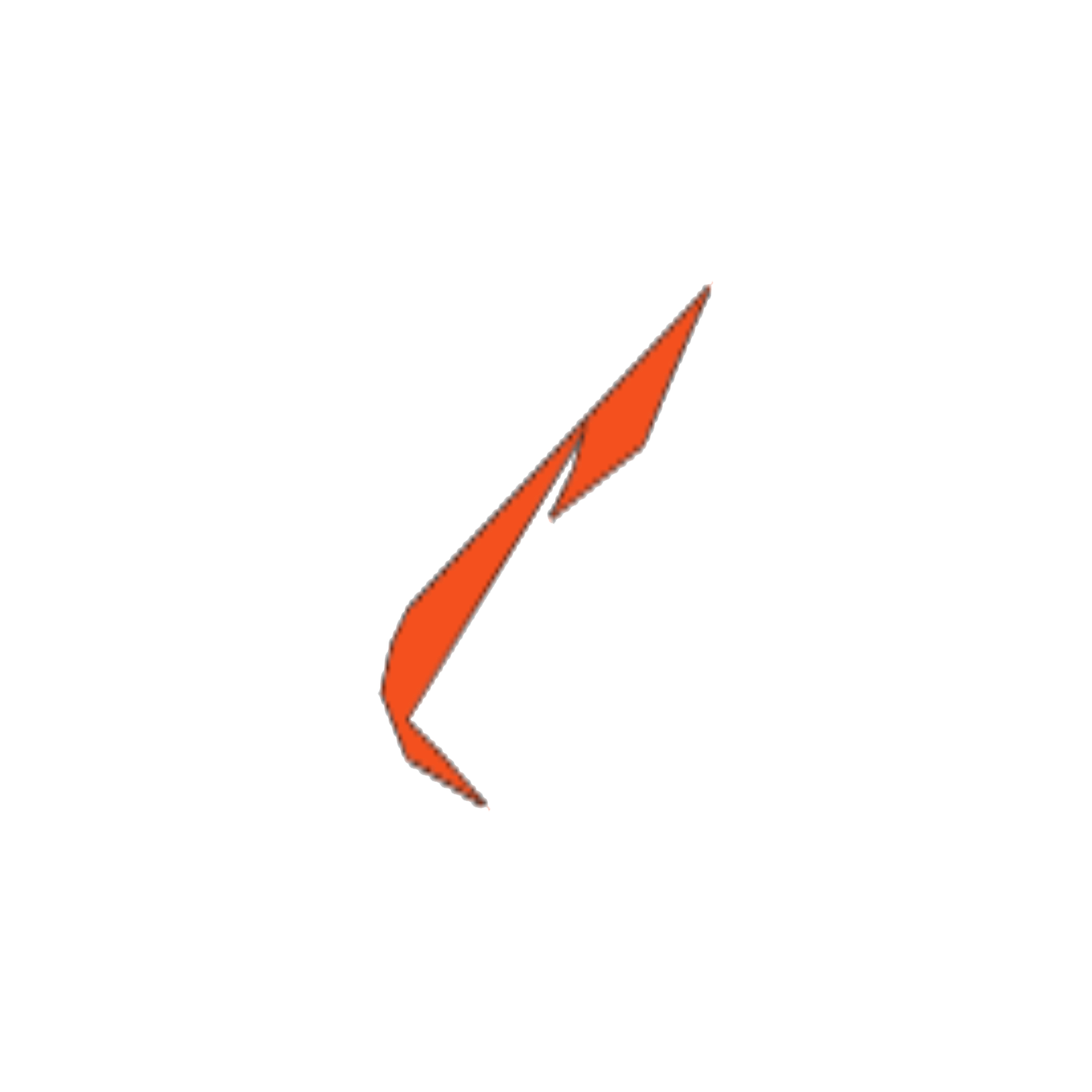} \\
    Gesture Output  & \includegraphics[width=1.5cm]{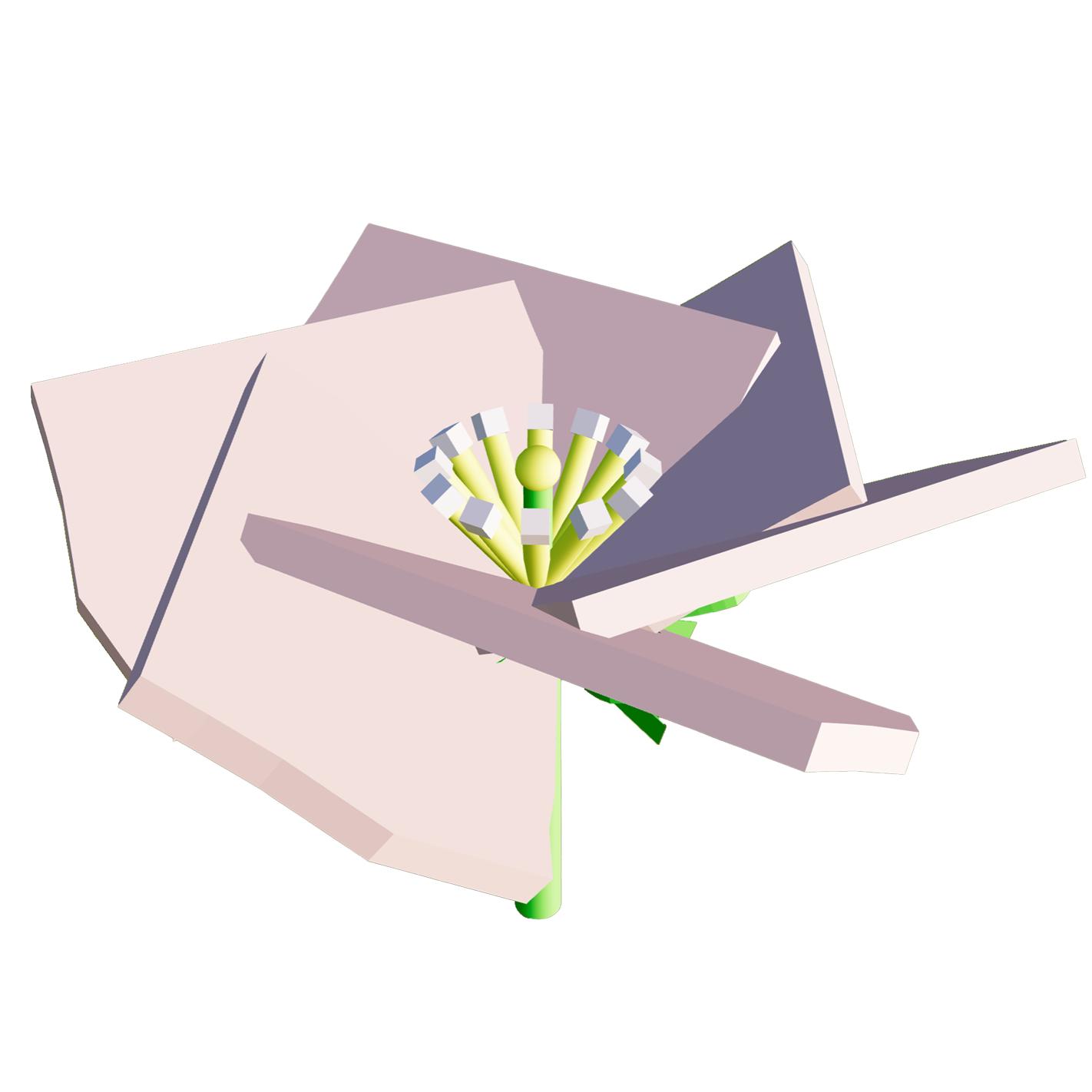}& \includegraphics[width=1.5cm]{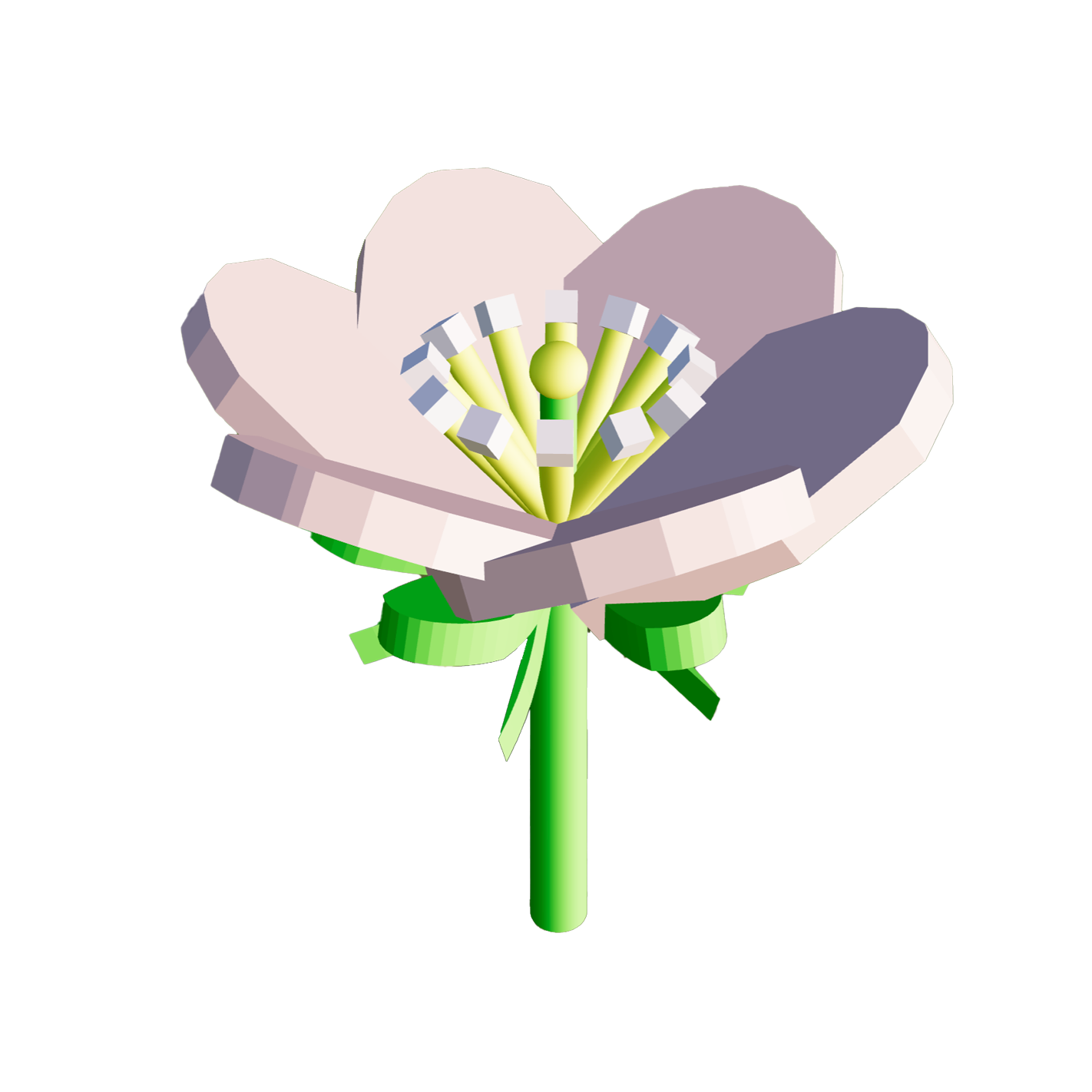}& \includegraphics[width=1.5cm]{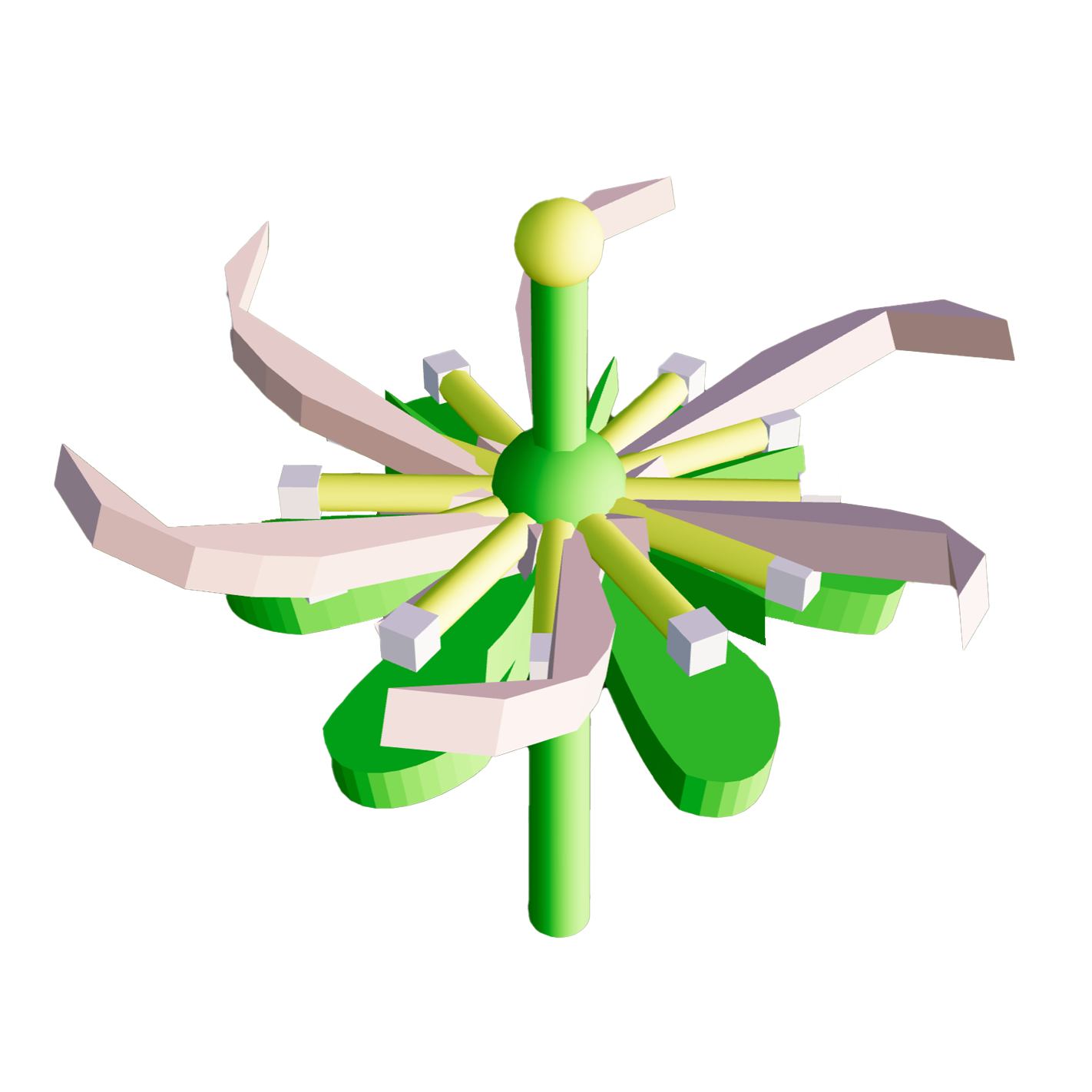} \\
    \bottomrule
  \end{tabular}
\end{table}

\begin{table}
   \caption{Position generation and modification}
  \label{tab:image}
  \begin{tabular}{>{\centering\arraybackslash}m{0.8in} *3{>{\centering\arraybackslash}m{0.65in}} @{}m{0pt}@{}}
    \toprule
    I/O & Case 1 & Case 2 & Case 3\\
        \midrule
    Voice Input & "47 degrees." & "Blooms a little bit." & "As open as my mind." \\
    Voice Output & \includegraphics[width=1.4cm]{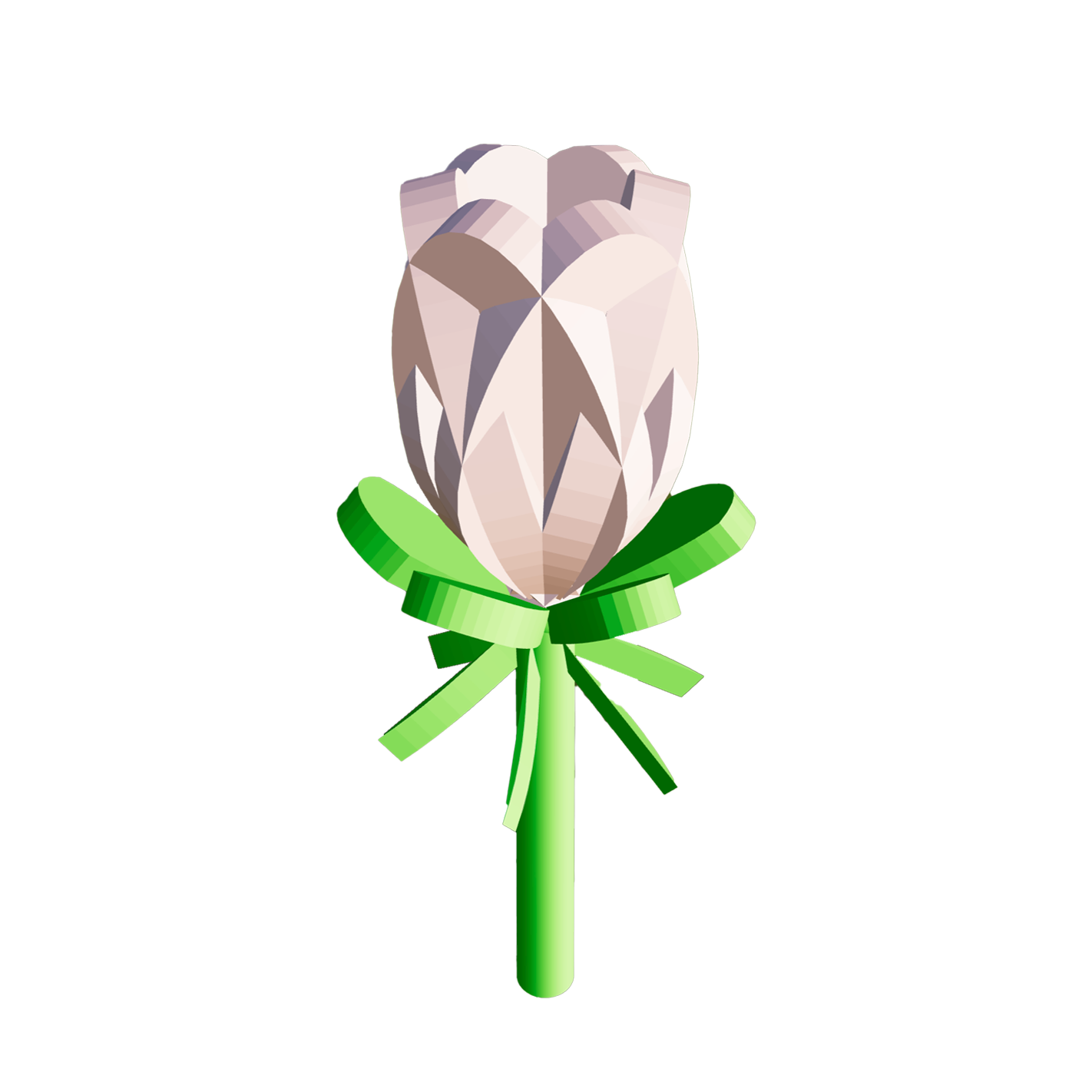}& \includegraphics[width=1.4cm]{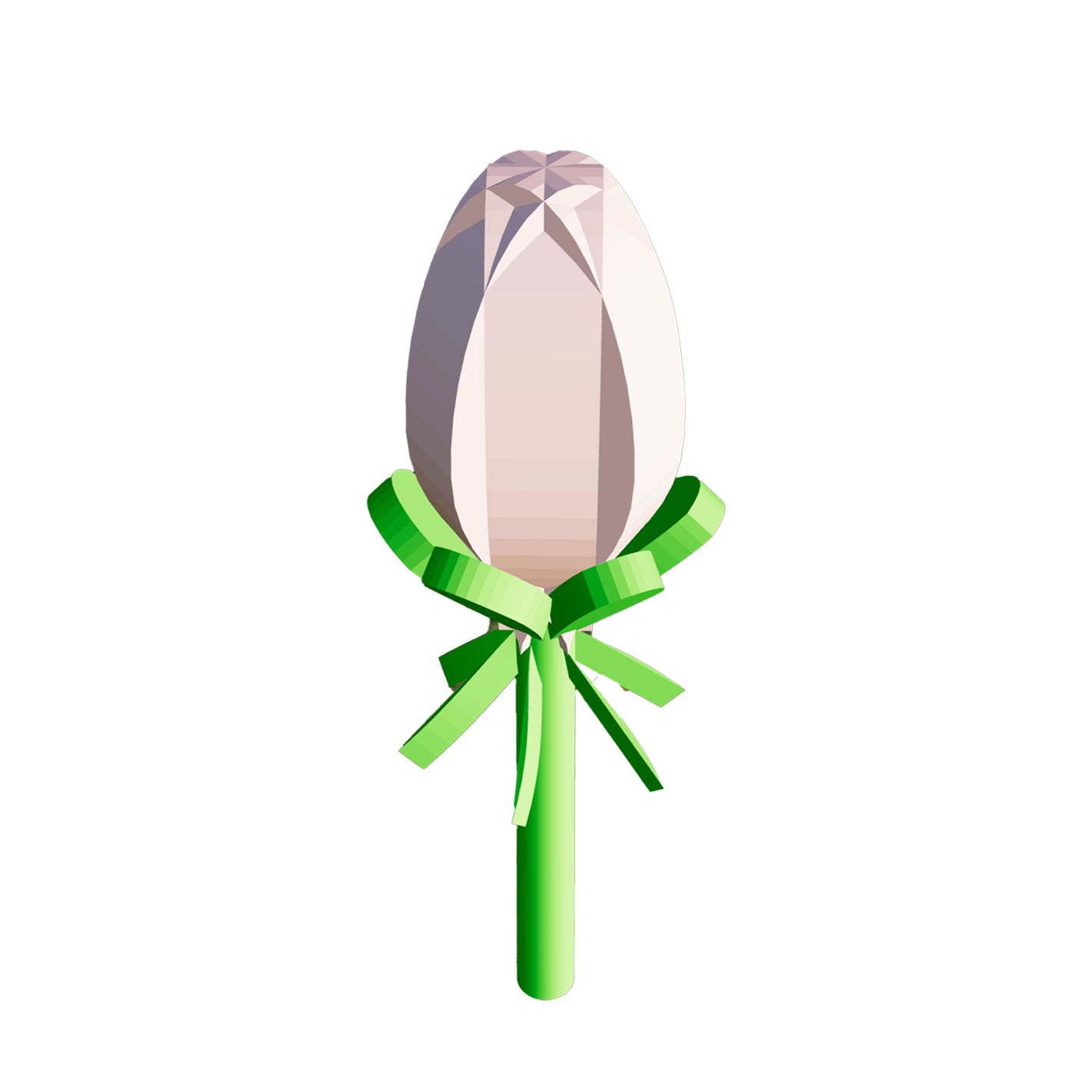}& \includegraphics[width=1.4cm]{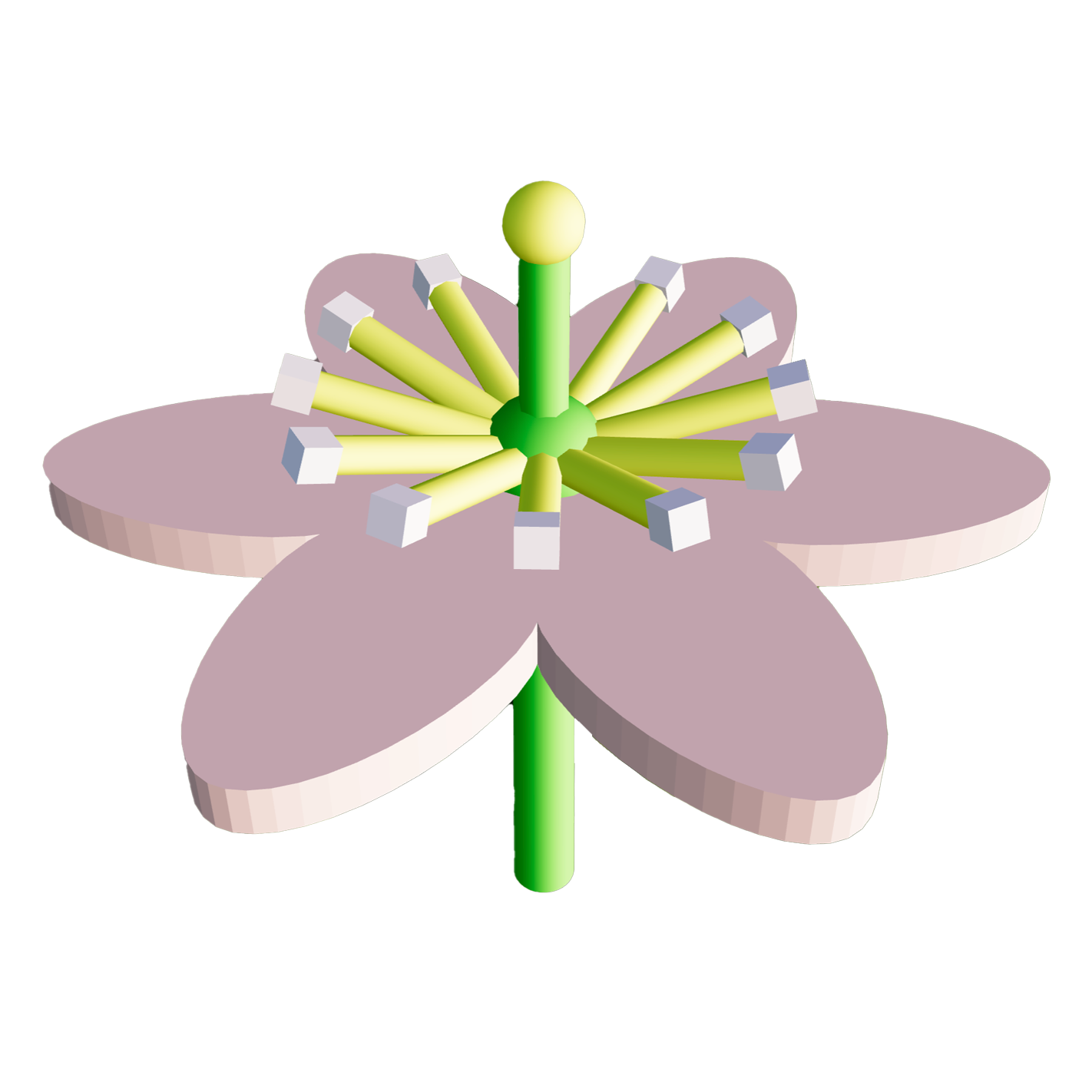} \\
    Gesture Input& \includegraphics[width=1.2cm]{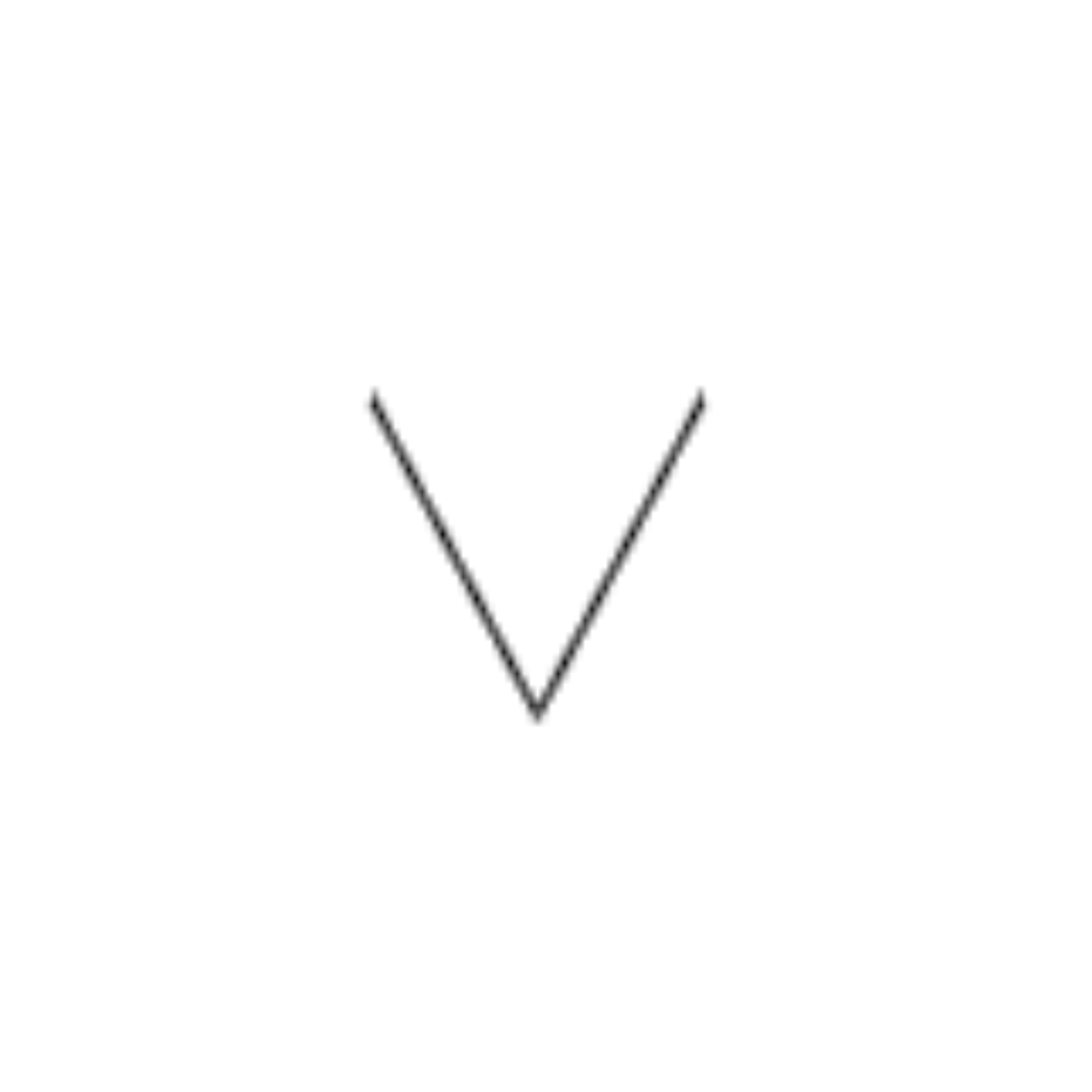}& \includegraphics[width=1.2cm]{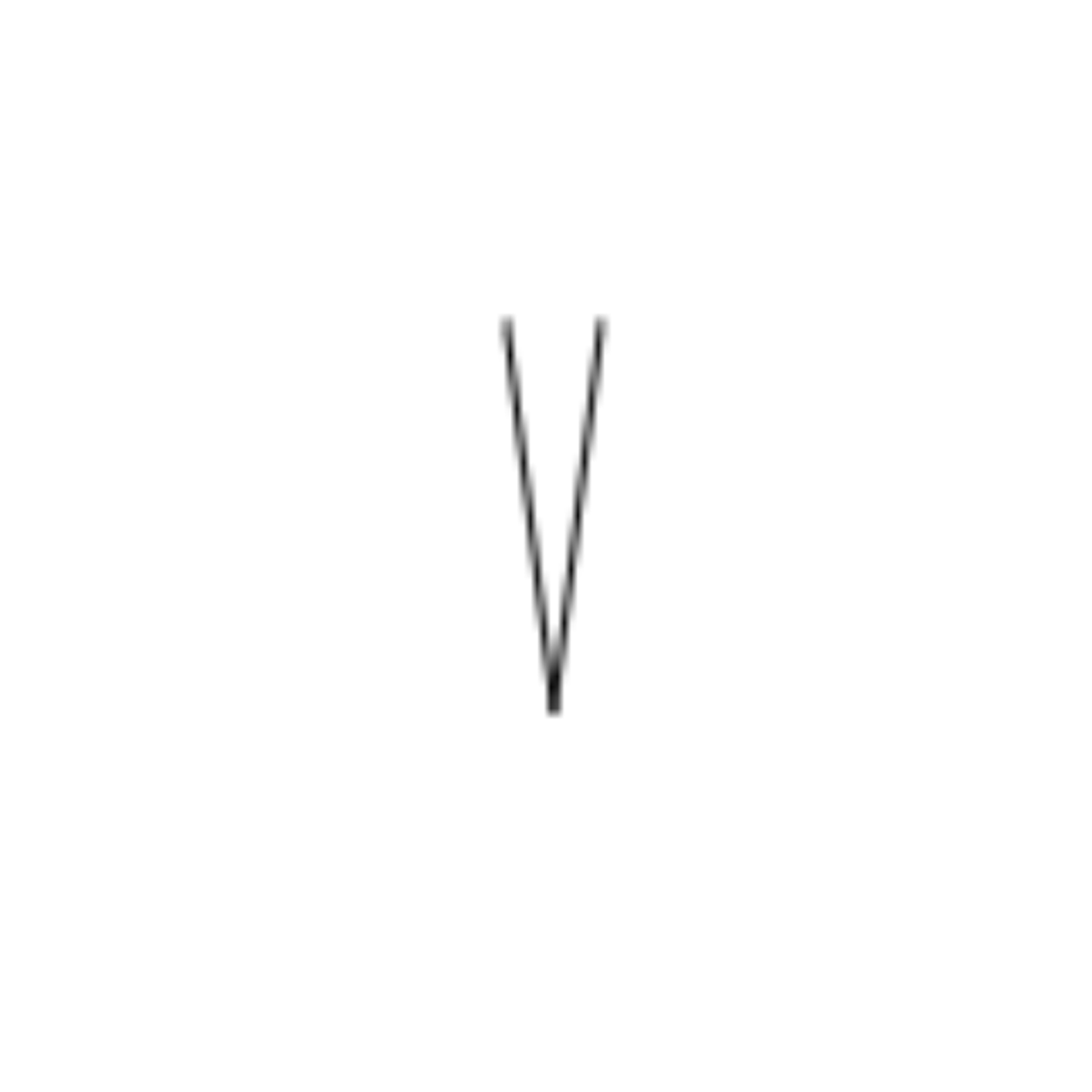}& \includegraphics[width=1.2cm]{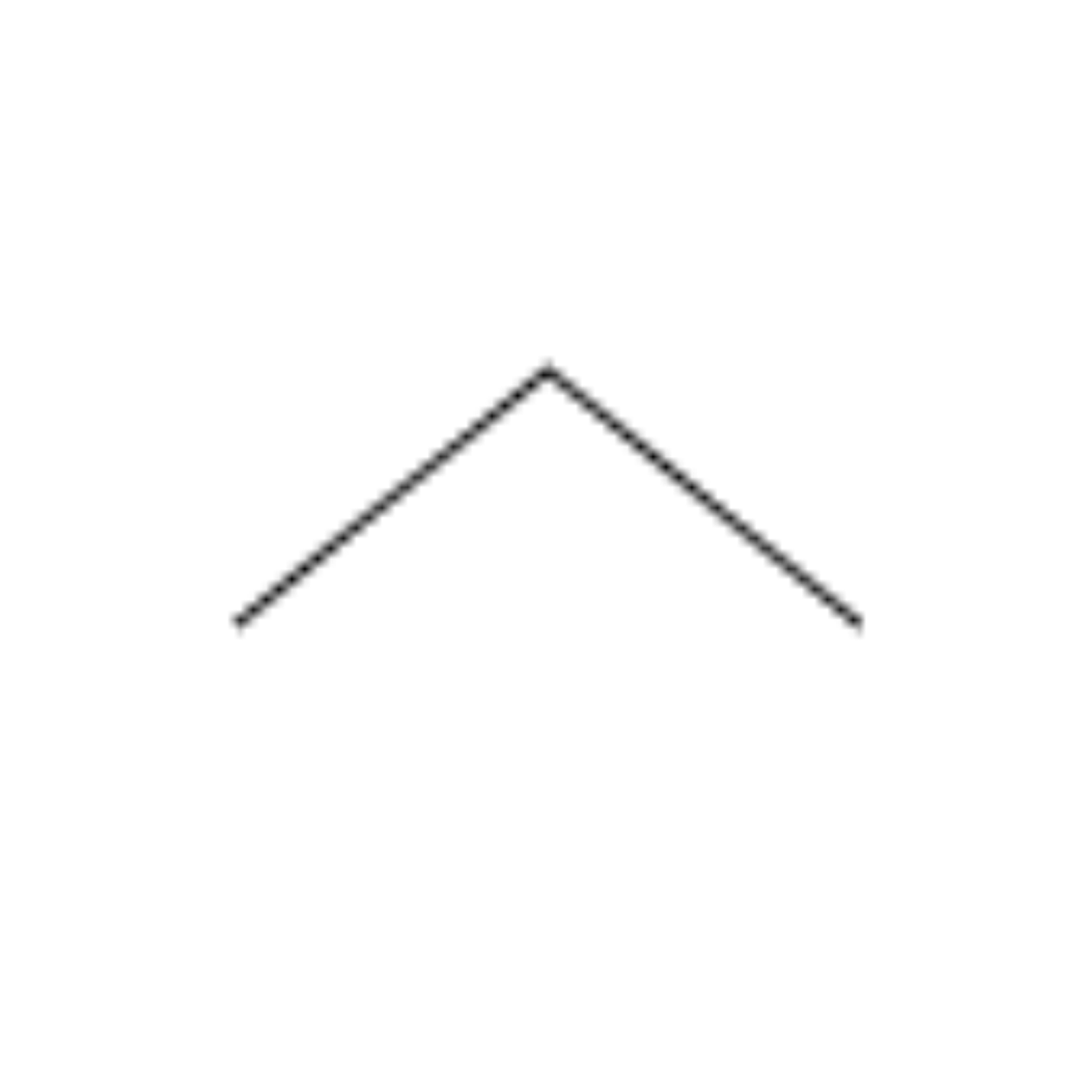} \\
    Gesture Output  & \includegraphics[width=1.4cm]{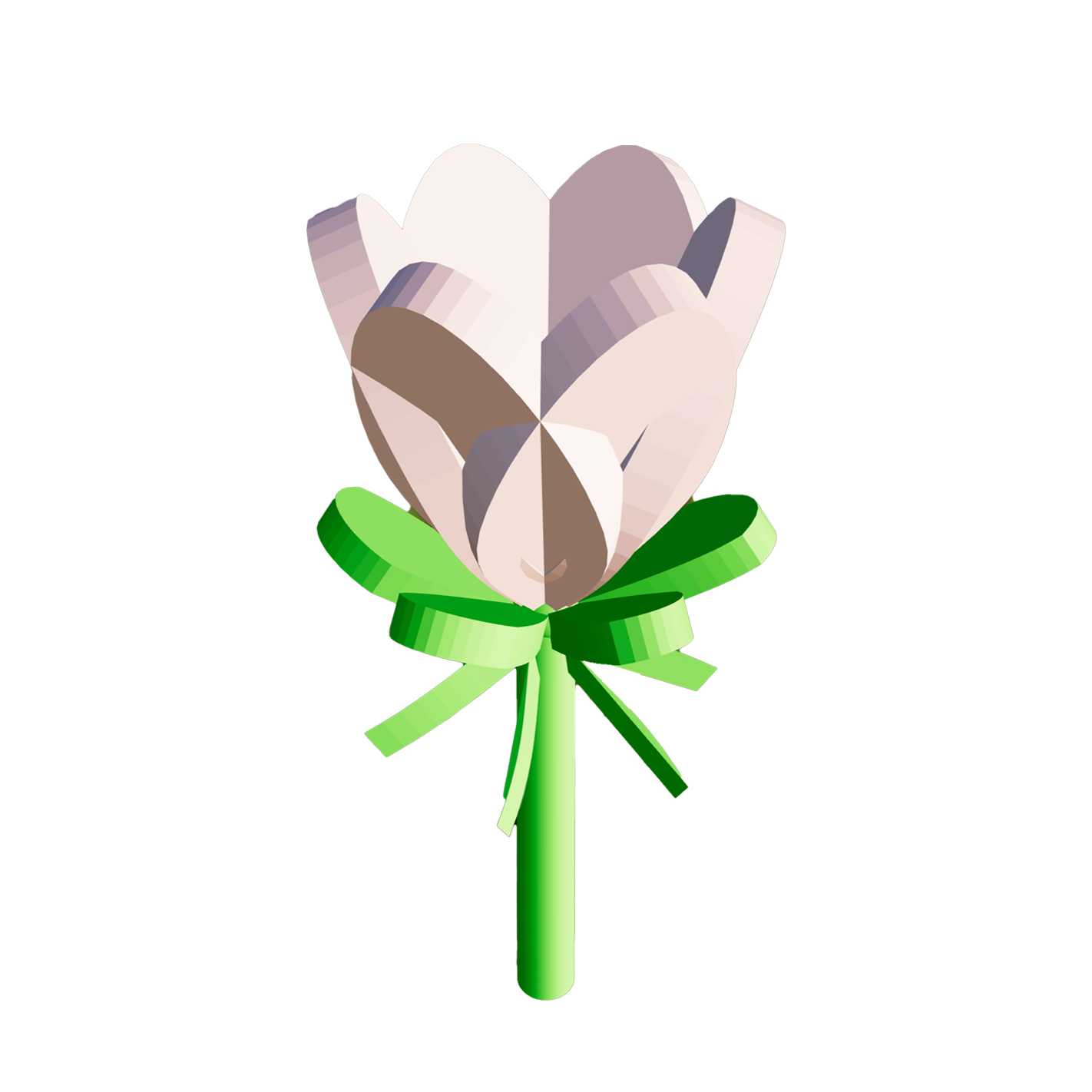}& \includegraphics[width=1.4cm]{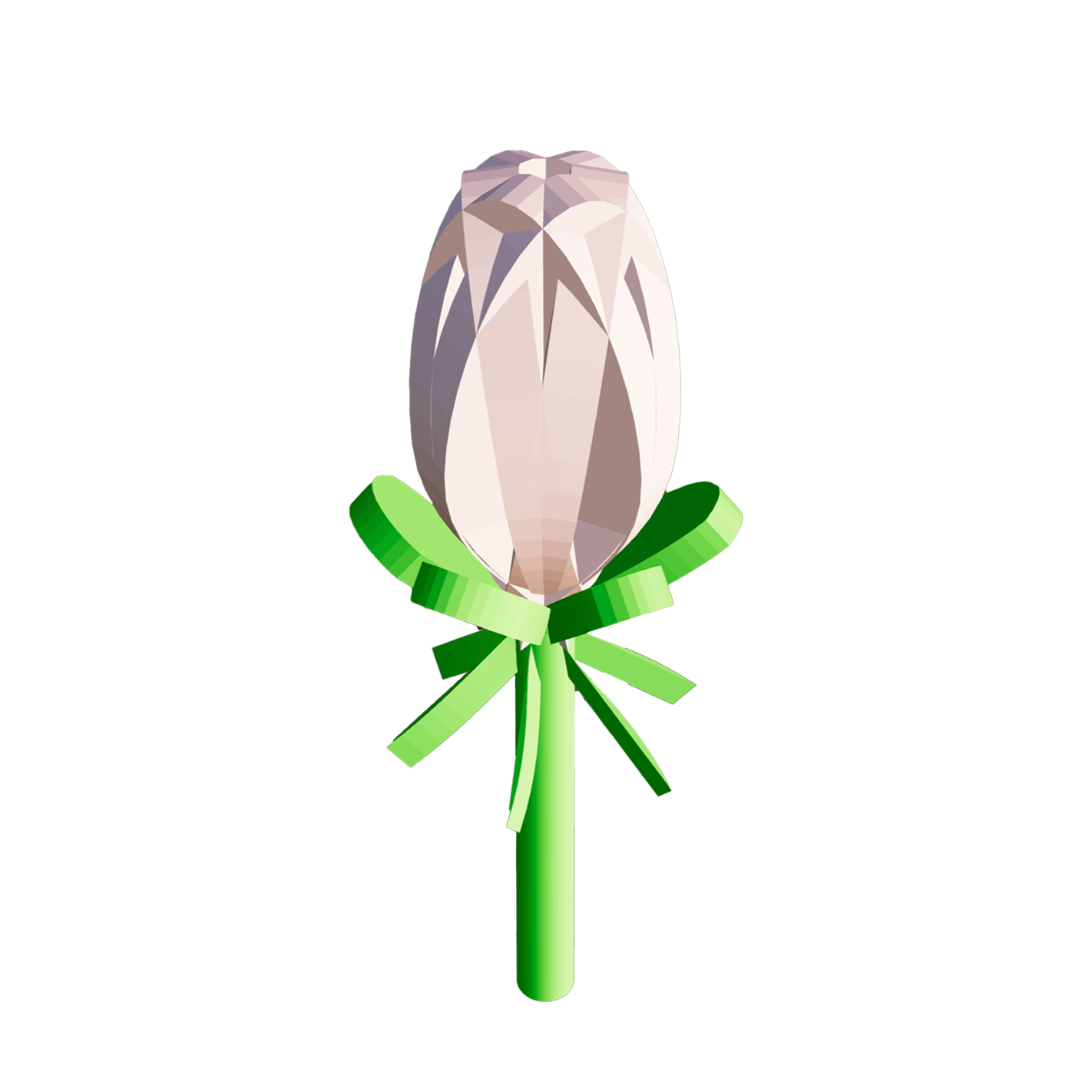}& \includegraphics[width=1.4cm]{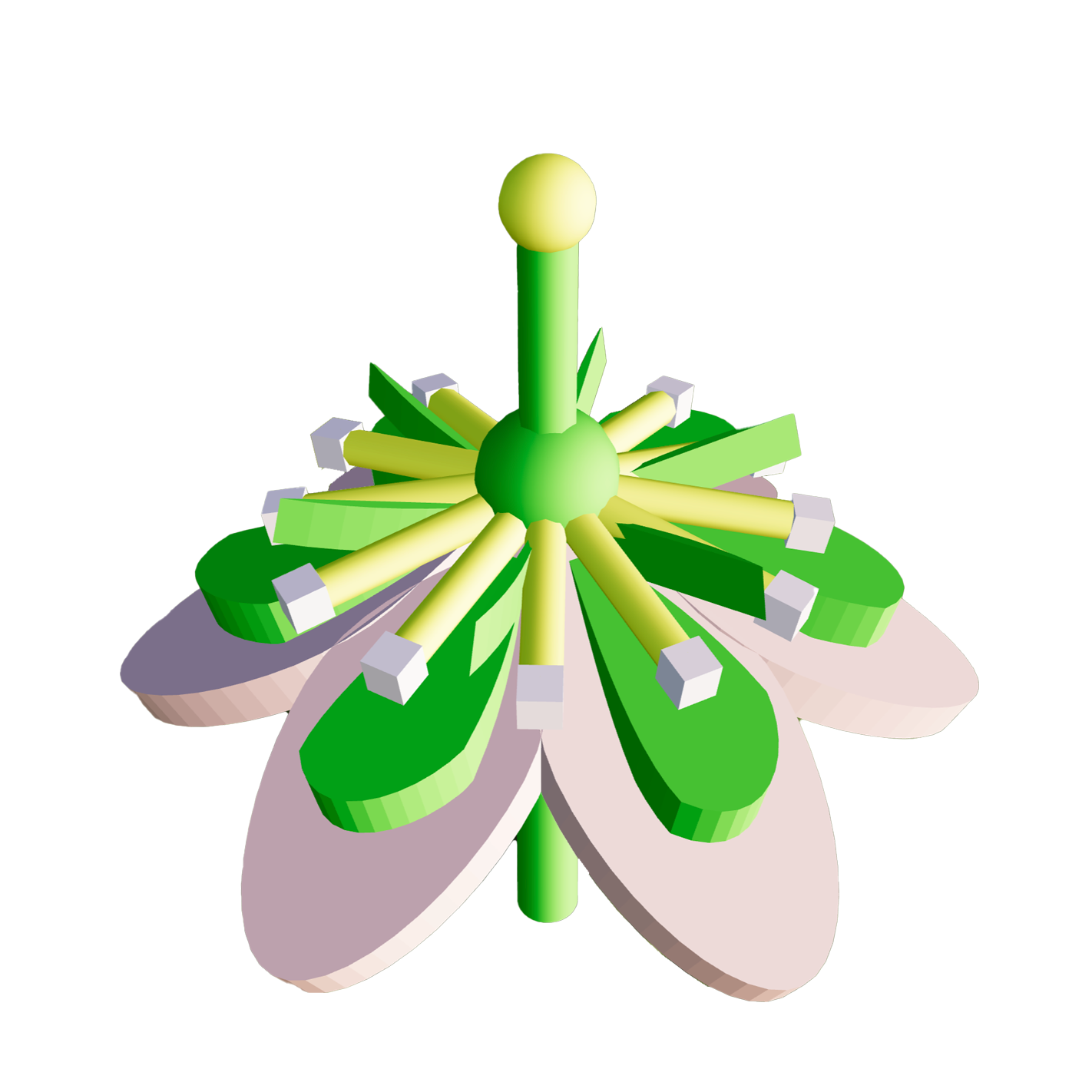} \\
    \bottomrule
  \end{tabular}
\end{table}

\begin{table}
   \caption{Color generation}
  \label{tab:image}
  \begin{tabular}{>{\centering\arraybackslash}m{0.8in} *3{>{\centering\arraybackslash}m{0.65in}} @{}m{0pt}@{}}
    \toprule
    I/O & Case 1 & Case 2 & Case 3\\
    \midrule
    Voice Input & "Standard HTML aqua." & "Hotter than fire." & "Reminds me of the courage of mankind."\\
    Voice Output & \includegraphics[width=1.5cm]{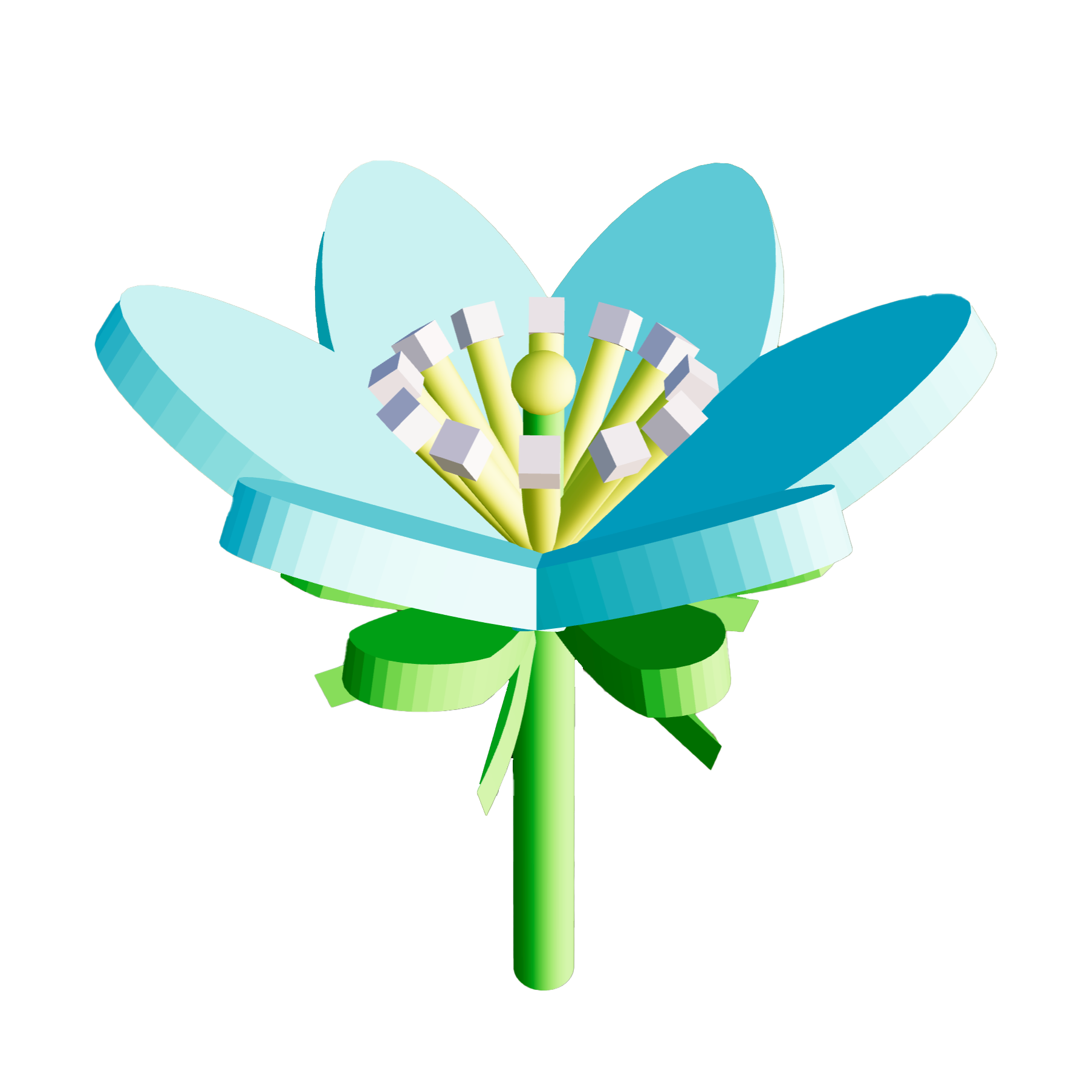}& \includegraphics[width=1.5cm]{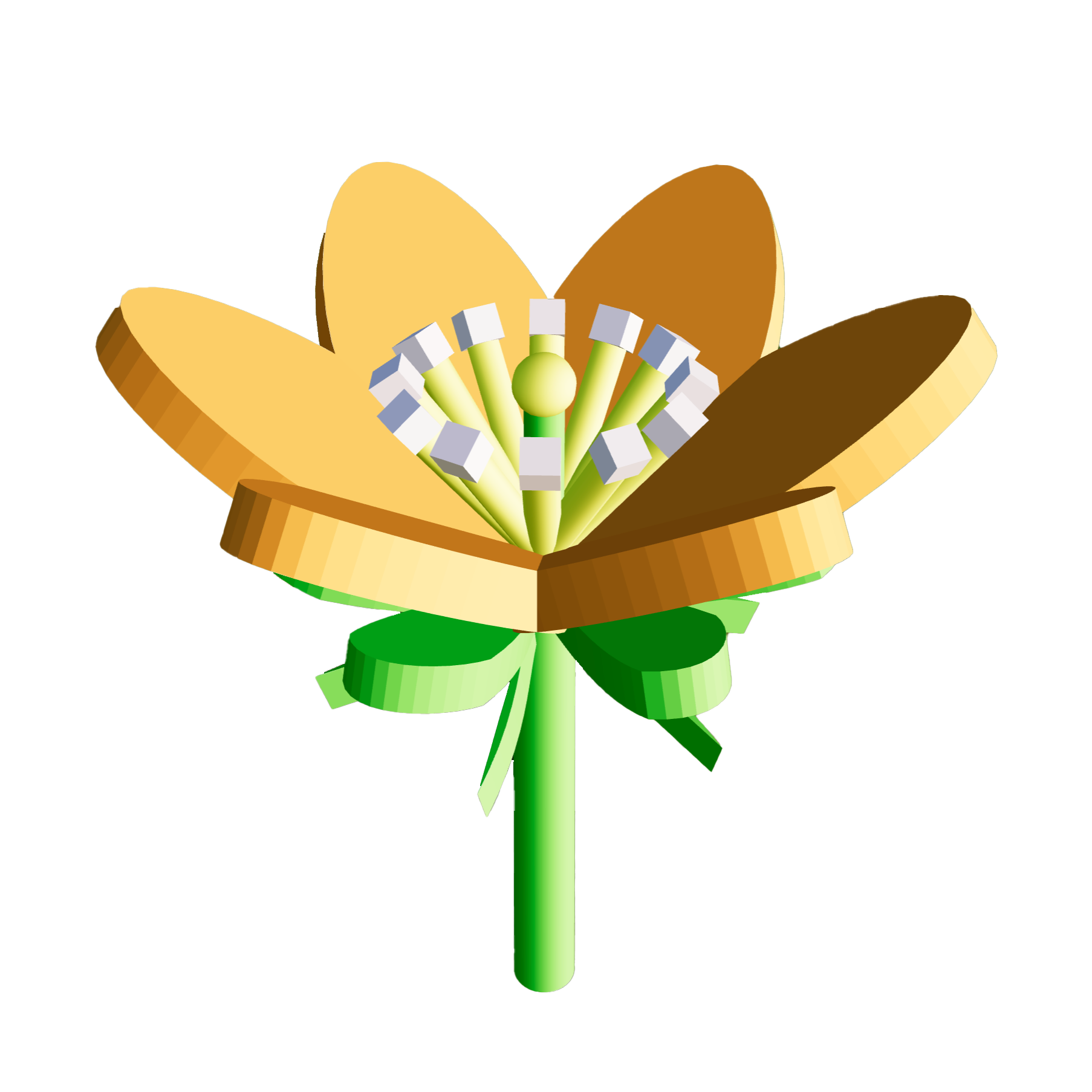}& \includegraphics[width=1.5cm]{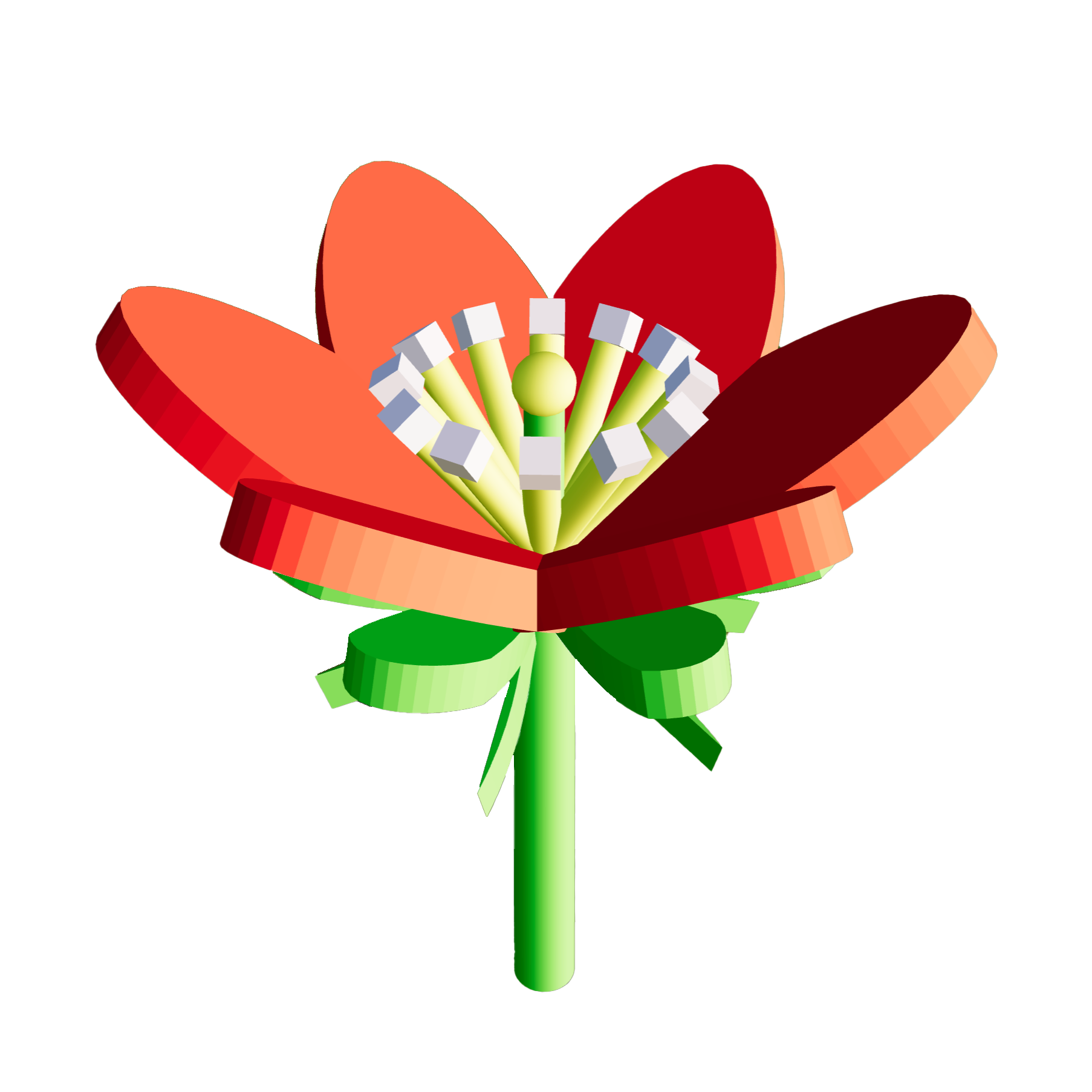} \\
    \bottomrule
  \end{tabular}
\end{table}

\subsection{Model modification with gesture}

While 3Description enables the direct generation of models from user voice input, here are scenarios where users may want to fine-tune the model by adjusting its position, shape, size, and other attributes. However, non-expert users often face difficulties in articulating precise details or providing quantitative feedback, such as defining the curvature using vectors. To address this challenge, 3Description provides users with an intuitive and easy-to-understand gesture-based solution to adjust the model, ensuring a more user-friendly experience.

By leveraging MediaPipe's gesture recognition technology, 3Description utilizes a total of 14 points as input from both hands, including the thumb and index finger. Then some of them are selected based on the user's chosen adjustment methods with different approaches:

(1) When users aim to approximate the general shape of the desired model using gestures, the whole 14-point position data is combined with a predefined prompt. This combined information is then converted into a generic language description. Subsequently, ChatGPT uses this description to generate a code block based on the Three.js format. Finally, the generated code block replaces the existing code to modify the model accordingly.

(2) In cases where users need to adjust specific dimensions or sizes of the model, 3Description selects the points that represent the tip of the index finger and the tip of the thumb. The distance between these points is calculated and treated as the length. Users have the flexibility to decide the unit of measurement for this length.

These approaches allow users to intuitively and effectively adjust the model based on their specific requirements.

\section{Discussion}

Based on user feedback and additional research, there are three key points that can be further explored:

\subsection{More ways of communication}

For instance, when it comes to the task of adjusting the angle of model components, users can simulate the angle change through gestures, which is captured and calculated by 3Description. However, during user testing, they expressed they could not intuitively confirm that the current value was what they desired, resulting in the model angle fluctuating in a small range. To solve this, a language detection module could be implemented to confirm the desired operation and obtain the final model angle. Additionally, facial recognition technology, such as detecting frowns to indicate dissatisfaction and smiles to indicate satisfaction, could be another possible solution to explore.

\subsection{More models and shapes}

Currently, the prompt instructs ChatGPT to create a 2D graphic plane which is later extruded to 3D. As an alternative to ChatGPT for model code generation, another approach to consider is training a dedicated 3D model generation network. This specialized network can be tailored to cater to a broader range of model categories, allowing for more precise and specific model code generation.

After model generation, users might also want to fine-tune specific components of the model. In the preliminary experiments, POINT-E was utilized to separate the model and convert them into individual meshes. This approach proved beneficial for adjusting the positional relationships between the components, in addition to enabling users to adjust the shape of individual components and eventually support the generation of complex shapes.

\subsection{More use cases}

3Description’s real-time model generation capability can be seamlessly integrated into video conferencing platforms. This integration gives visual feedback directly from participants' descriptions, transforming abstract ideas into tangible 3D models. Consequently, the inclusion of explicit 3D models enhances meeting efficiency by facilitating discussions based on real-time visual presentation.

\section{Conclusion}

In this study, the researcher examined the prevailing 3D modeling approaches to identify the pain points experienced by non-professionals. To gain a comprehensive understanding of how non-professionals work on 3D modeling, the researcher interviewed users from non-art or design backgrounds to capture and analyze their descriptions of their 3D modeling needs. Taking into account the users' description preferences, the researcher decided to allow users to describe their models through voice and gesture. Using OpenAI's Whisper and Google's MediaPipe technology, the researcher successfully captured users' verbal input and gestures. The collected input was then processed, incorporating specific prompts, and fed into ChatGPT to obtain interchangeable code blocks to achieve real-time model generation and modification. By combining state-of-the-art technologies, 3Description offers a convenient and efficient 3D modeling solution for non-professionals, enabling them to stay up-to-date with the latest trends and actively participate in the field of 3D modeling.

In the future, acquiring annotated datasets and training dedicated AI models for 3Description will enhance multimodal human-AI 3D collaboration and promote better alignment between 3D models and natural human descriptions.

\begin{acks}
This paper is a part of the author's graduate thesis in the Interactive Telecommunications Program (ITP) at New York University Tisch School of the Arts. Special thanks to Sharon De La Cruz as thesis advisor, the ITP community, and Yuan Li for technical discussions.
\end{acks}

\bibliographystyle{ACM-Reference-Format}
\bibliography{sample-base}

\end{document}